\documentclass[preprint,floatfix,showkeys]{revtex4-2}

\usepackage[english]{babel}

\usepackage[letterpaper,top=2cm,bottom=2cm,left=3cm,right=3cm,marginparwidth=1.75cm]{geometry}

\usepackage{amsmath}
\usepackage{graphicx}
\usepackage{natbib}
\usepackage[colorlinks=true, allcolors=blue]{hyperref}

\begin{document}
\title{Understanding following patterns among high-performance athletes}
\author{Jorge P. Rodríguez}
\email{jorgeprodriguezg@gmail.com}
\affiliation{Instituto de Física Interdisciplinar y Sistemas Complejos (IFISC), CSIC-UIB, E-07122 Palma de Mallorca, Spain}
\affiliation{C.A. Illes Balears, Universidad Nacional de Educación a Distancia (UNED), 07009 Palma de Mallorca, Spain}
\author{Lluís Arola-Fernández}
\affiliation{Instituto de Física Interdisciplinar y Sistemas Complejos (IFISC), CSIC-UIB, E-07122 Palma de Mallorca, Spain}

\begin{abstract}
Professional sports enhance interaction among athletes through training groups, sponsored events and competitions. Among these, the Olympic Games represent the largest competition with a global impact, providing the participants with a unique opportunity for interaction. We studied the following patterns among highly successful athletes to understand the structure of their interactions. We used the list of Olympic medallists in the Tokyo 2020 Games to extract their follower-followee network in Twitter, finding 7,326 connections among 964 athletes. The network displayed frequent connections to similar peers in terms of their features including sex, country and sport. We quantified the influence of these features in the followees choice through a gravity approach capturing the number of connections between homogeneous groups. Our research remarks the importance of datasets built from public exposure of professional athletes, serving as a proxy to investigate interesting aspects of many complex socio-cultural systems at different scales.

\end{abstract}

\keywords{Complex networks, social networks, directed networks, Olympic Games, assortativity, gravity model}
\maketitle
\section{Introduction}

The structure of interactions frequently helps understand social dynamics. Such structure is typically represented as complex networks \cite{boccaletti2006complex}, which are often characterized by heterogeneity where a few individuals, known as hubs, can concentrate a large fraction of the total number of interactions \cite{barabasi1999emergence,broido2019scale}. Although cognitive constraints limit the amount of interactions that a person can sustain, for example through the Dunbar numbers \cite{tamarit2018cognitive}, new technologies allow individuals to connect to a larger number of people. This is particularly interesting in the case of directed social networks, where interactions do not have to be reciprocal, and thus an individual can receive information from other users that may not even know her. Such platforms drove the emergence of \emph{influencers}, whose  profiles in online social networks are massively followed, and consequently institutions and companies frequently choose them for their advertising campaigns. People placed in strategic network locations, known as influential spreaders \cite{kitsak2010identification,chen2012identifying} in the terminology of complex networks, play a key role in the outcome of the global network dynamics \cite{pei2013spreading}. The driving force of influential spreaders often coexists with the presence of strong assortative (\emph{i.e.}, homophilic) patterns in the social networks \cite{mcpherson2001birds}, observed when users tend to interact more with people with shared preferences or that are similar to them, which can lead to disadvantages for the minorities \cite{karimi2018homophily}. These assortative patterns also shape global network dynamics \cite{yavacs2014impact,steinegger2022groups,burgio2022homophily}, so characterizing them becomes another key factor to understand how information spreads through networks in order to design viral advertising strategies \cite{sheikhahmadi2017identification}.

High-performance professional athletes are an example of users in social networks that influence general population's opinion and activities \cite{pegoraro2010twitter}. The most popular are able to reach global audiences, becoming reference models for many people, especially children, and attractive targets for commercial advertising \cite{anagnostopoulos2018advertising}. In online social media, these sport celebrities also engage in interactions with their followers \cite{gibbs2014engaging}, potentially impacting, with their opinions and activities, some aspects of the cultural evolution at a global scale. Beyond online platforms, the Olympic Games represent the major sport gathering in the world. Except for a few official supporters, other advertising campaigns are not allowed during the event \footnote{\url{https://olympics.com/ioc/documents/international-olympic-committee}}. However, the broad relevance of this competition has allowed athletes to impact the spreading of relevant topics such as equality, democracy, peace or job conditions throughout history \cite{tomlinson2005history}. 

The significant impact of world-class athletes enrolled in the Olympic Games and their public exposure in online social platforms offers a great opportunity to understand and quantify how information flows among these highly influential spreaders. A microscopic analysis of the interaction patterns among individual athletes can be complemented by a more macro-scale view, where interaction can be analysed through groups of athletes sharing features such as nationality, sport or sex, aiming to reveal assortativity fingerprints among groups. In this work, we explore for the first time the empirical network of following patterns on Twitter among the most influential athletes, the Olympic medallists, by analysing the presence of super-influential spreaders and homophily following patterns (assortativity) in the network, and characterising cross-group interactions via a gravity model approach.

\section{Results}

% Fig. 1. Network
\begin{figure*}[hbt]
\includegraphics[width=\textwidth]{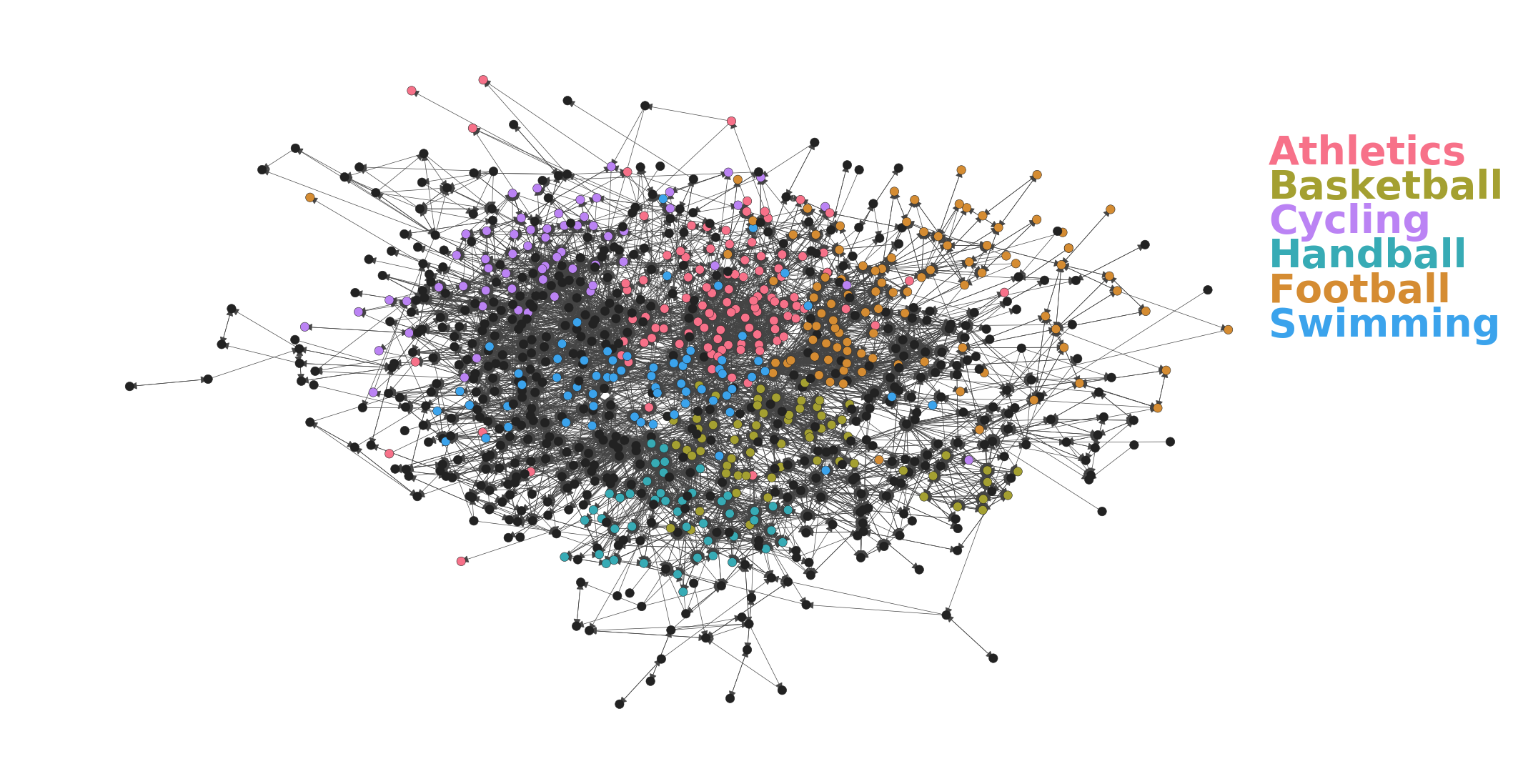}
\caption{{\bf Following network between Olympic medallists.} Arrows in links point from the followers to the followees and the node colours represent medallists in sports displaying a meso-scale structure, including athletics (red), basketball (olive), cycling (violet), handball (green), football (orange), swimming (blue). Athletes practising other sports are represented in black. }
\label{figs:fig1}
\end{figure*}

\subsection{A directed following network}

We created a database composed of 1,052 medallists (from 76 countries in 36 sports) in the Tokyo 2020 Olympic Games who had a public Twitter account. The use of this database to query the Twitter API led to the discovery of $E=7326$ following links (\emph{i.e.} connecting a follower to a followee) between $N=964$ different individuals (representing 64 countries in 36 sports, including both sexes). These links accounted for 2.4\% of the total number of these athletes' followees in Twitter (\emph{i.e.}, the remaining 97.6\% of links including these athletes as followers were directed towards users that were not included in our database). Interestingly, although the nodes in the network represented specialized individuals, the network was sparse, as the links that we observed represent only 0.79\% of the possible $N(N-1)$ connections.

The network displayed a largest connected component, defined as the largest component of nodes connected by links (ignoring their directness), of size 956 nodes, while the strongly connected largest component, which restricts the largest component to the set of nodes displaying a return path following the directed network between all possible pairs \cite{dorogovtsev2001giant}, had a size of 827 nodes. The graphical representation of the network revealed the presence of meso-scale structures with high density of links associated to specific sports, including athletics, basketball, cycling, handball, football and swimming (Fig. \ref{figs:fig1}).

%Celebrities
\begin{table*}[hbt]
    \centering
    \begin{tabular}{|c|c|c|c|c|c|}
    \hline
    Rank & Name & Sport & Country & $k_{\text{in}}$ & Medal(s) \\ \hline
    1 & Kevin Durant & Basketball & United States & 69 & G \\ \hline
    2 & Allyson Felix & Athletics & United States & 56 & G, B \\ \hline
    3 & Teddy Riner & Judo & France & 54 & G, B \\ \hline
    4 & Alex Morgan & Football & United States & 52 & B \\ \hline
    5 & Simone Biles & Gymnastics & United States & 50 & S, B \\ \hline
    6 & Megan Rapinoe & Football & United States & 48 & B \\ \hline
    7 & Adam Peaty & Swimming & Great Britain & 46 & G, G, S \\ \hline
    8 & Nikola Karabatic & Handball & France & 43 & G \\ \hline
    9 & Noah Lyles & Athletics & United States & 40 & B \\ \hline
    9 & Tom Daley & Diving & Great Britain & 40 & G, B \\ \hline
    \end{tabular}
    \caption{{\bf Rank of the 10 most followed athletes.} Athletes are sorted by decreasing in-degree $k_{\text{in}}$. The considered medals are gold (G), silver (S) and bronze (B) representing, respectively, the 1$^{\text{st}}$, 2$^{\text{nd}}$ and 3$^{\text{rd}}$ positions in the podium.}
    \label{tab:t1}
\end{table*}

The average degree of the network was $\langle k \rangle = E/N =7.60$. For directed networks, this represents the total number of followers (as well as the total number of followees), divided by the number of individuals included in the network. However, the values for individual $i$ of the in-degree $k_{i,\text{in}}$ (number of followers) and out-degree $k_{i,\text{out}}$ (number of followees), varied significantly from $\langle k \rangle$, displaying broad distributions, with 32\% of the individuals with the highest $k_{\text{in}}$ being the targets (\emph{i.e.}, followees) of 68\% of the total number of links, and 33\% of the individuals with the highest $k_{\text{out}}$ being the source (\emph{i.e.}, followers) of 67\% of the links. We used the in-degree to extract the 10 most popular athletes in the network, finding both well-known athletes (for example, Kevin Durant, Simone Biles or Megan Rapinoe), and interestingly also other athletes with high $k_{\text{in}}$ who are not so popular among the general public (Table \ref{tab:t1}). Specially, we observed a majority of athletes from the United States (US), with 6 individuals included in this top-10 ranking.

%Reciprocity
The links in our network were directed, as the connections on Twitter are not reciprocal, meaning that user A following user B does not imply that B follows A. Considering the importance of this feature, we quantified it, finding that, out of 4898 unique undirected links, 2428 (49.6\%) were reciprocal (\emph{i.e.}, almost half of the connections were two-way). This generic pattern displayed remarkable differences at individual level. We used the Jaccard index, a standard way to quantify the overlap between different sets, to estimate the reciprocity for a given individual $i$, obtaining $J_i$ as the intersection divided by the union of the set of her followers and the set of her followees. Considering $a_{ij}=1$ when $i$ followed $j$ and 0 otherwise, the Jaccard index was defined as
\begin{equation}
J_i=\frac{\sum_j a_{ij} a_{ji}}{\sum_j(a_{ij}+a_{ji}-a_{ij}a_{ji})}
\label{jacc0}
\end{equation}

Although the Jaccard index has been broadly used, in the context of our directed network we observed that a small value of $J_i$ could be the result of different scenarios, including: a) similar in and out-degrees, but low intercept among the specific followers and followees, b) $k_{i,\text{in}} \gg  k_{i,\text{out}}$, or c) $k_{i,\text{in}} \ll  k_{i,\text{out}}$. To distinguish between such scenarios, we introduced the in-reciprocity $r_{i,\text{in}}$ as the fraction of $i$'s followees that followed her back (\emph{i.e.}, the potential of $i$ to get a follow back when following a new user), and the out-reciprocity $r_{i,\text{out}}$ as the fraction of $i$'s followers that $i$ followed. Mathematically, the in- and out-reciprocities were given by:
\begin{align}
r_{i,\text{in}} = \frac{\sum_j a_{ij} a_{ji}}{\sum_j a_{ij}} \\
r_{i,\text{out}} = \frac{\sum_j a_{ij} a_{ji}}{\sum_j a_{ji}} 
\end{align}

\begin{figure}
\includegraphics[width=0.5\textwidth]{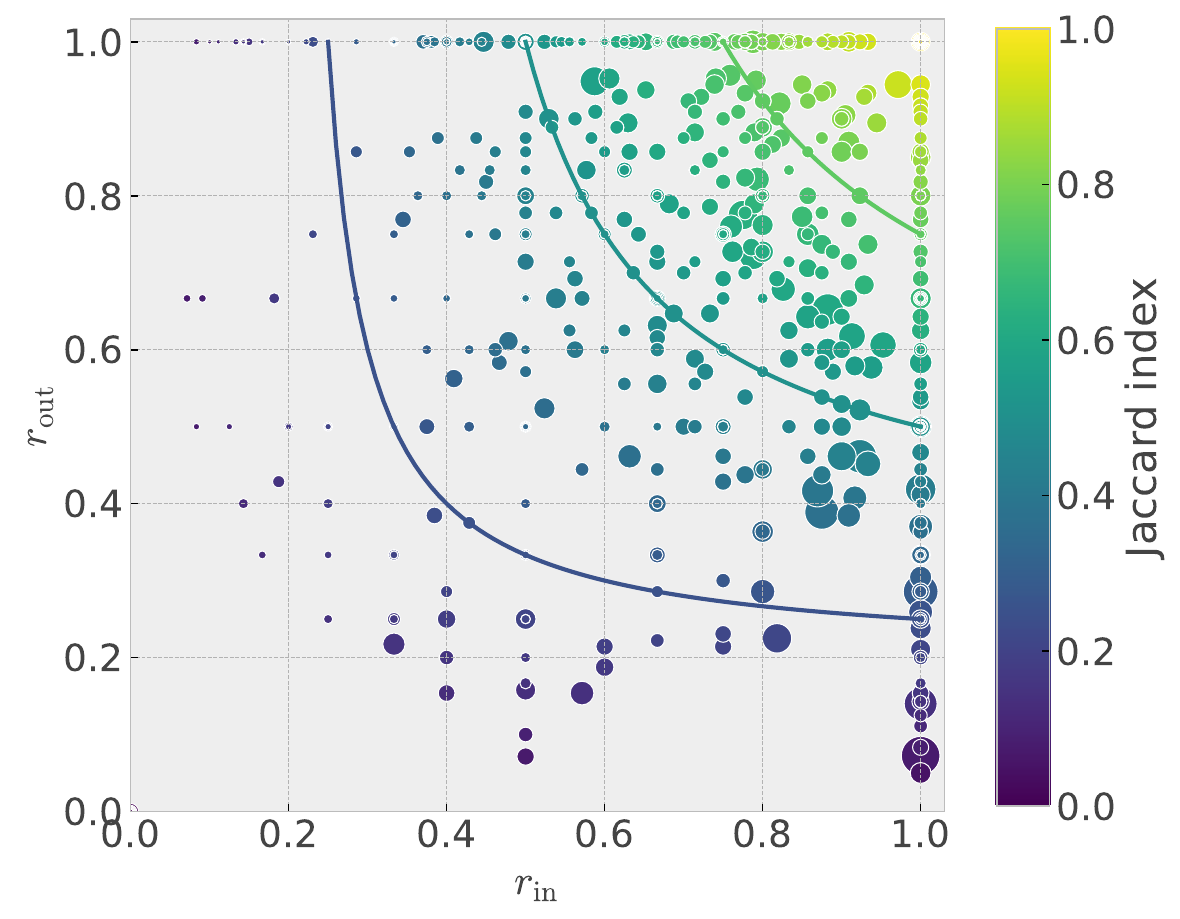}
\caption{{\bf Reciprocity patterns in the network.} For each ego node, we compute the in-reciprocity $r_{\text{in}}$ (fraction of ego's followees that follow her) and the out-reciprocity $r_{\text{out}}$ (fraction of ego's followers that she follows). Symbol sizes depict the node's in-degree (number of followers in the network) and the colors represent the Jaccard index $J$ between the set of followers and the set of followees of each node. We have represented all the possible combinations of $(r_{\text{in}},r_{\text{out}})$ on the solid lines for $J=\{0.25,0.5,0.75\}$, following Eq. (\ref{jaccrecipr}).}
\label{figs:fig2}
\end{figure}

In fact, the Jaccard index $J_i$ was related to $r_{i,\text{in}}$ and $r_{i,\text{out}}$ through the equation
\begin{equation}
J_i^{-1} = r_{i,\text{in}}^{-1} + r_{i,\text{out}}^{-1}-1
\label{jaccrecipr}
\end{equation}
that is valid for $J_i \neq 0$ and, considering that $r_{i,\text{in}},r_{i,\text{out}}\in [0,1]$, had a unique solution $r_{i,\text{in}}=r_{i,\text{out}}= 1$ for $J_i= 1$ (\emph{i.e.}, when the set of $i$'s followers exactly matched the set of her followees). However, for $J_i \neq 0,1$, the values of $r_{i,\text{in}}$ and $r_{i,\text{out}}$ were undetermined, meaning that the solution of Eq. (\ref{jaccrecipr}) was a curve in the $(r_{\text{in}},r_{\text{out}})$ space (see the solutions for $J=\{0.25,0.5,0.75\}$ represented as lines in Fig. \ref{figs:fig2}). Consequently, we decided to study the reciprocity in the $(r_{\text{in}},r_{\text{out}})$ space, depicting the Jaccard index as an orientative value in the colour of each individual data point. At individual level, we detected frequent observations of high values of the in-reciprocity and out-reciprocity (dots in Fig. \ref{figs:fig2}), but also a non-negligible amount of individuals that displayed an intermediate behaviour between both. Interestingly, representing the in-degree as the symbol size informed on the frequent behaviour of the most popular athletes, who displayed a high in-reciprocity and low out-reciprocity, with values, respectively, higher than 0.8 and lower than 0.45. Particularly, 4 out of the 10 most popular athletes (Table \ref{tab:t1}) displayed the maximum in-reciprocity (\emph{i.e.}, $r_{\text{in}}=1$), and $r_{\text{out}}<0.45$. However, this was not the case of all the athletes with high in-degree, as we observed two individuals with high out-reciprocity (larger than 0.9) and variable in-reciprocity. Specifically, $r_{\text{in}}$ was 0.54 for the individual with the highest number of followees in the network, \emph{i.e.} with the highest $k_{\text{out}}$ and the in-reciprocity was 0.97 for an individual with $k_{\text{in}}=k_{\text{out}}+1$ and a high overlap between her followers and followees).

% Fig. 2. As directed network, characterize reciprocity

\subsection{Quantifying assortative patterns}
% Fig. 3. Sex assortativity
\begin{figure}[hbt]
\includegraphics[width=0.5\textwidth]{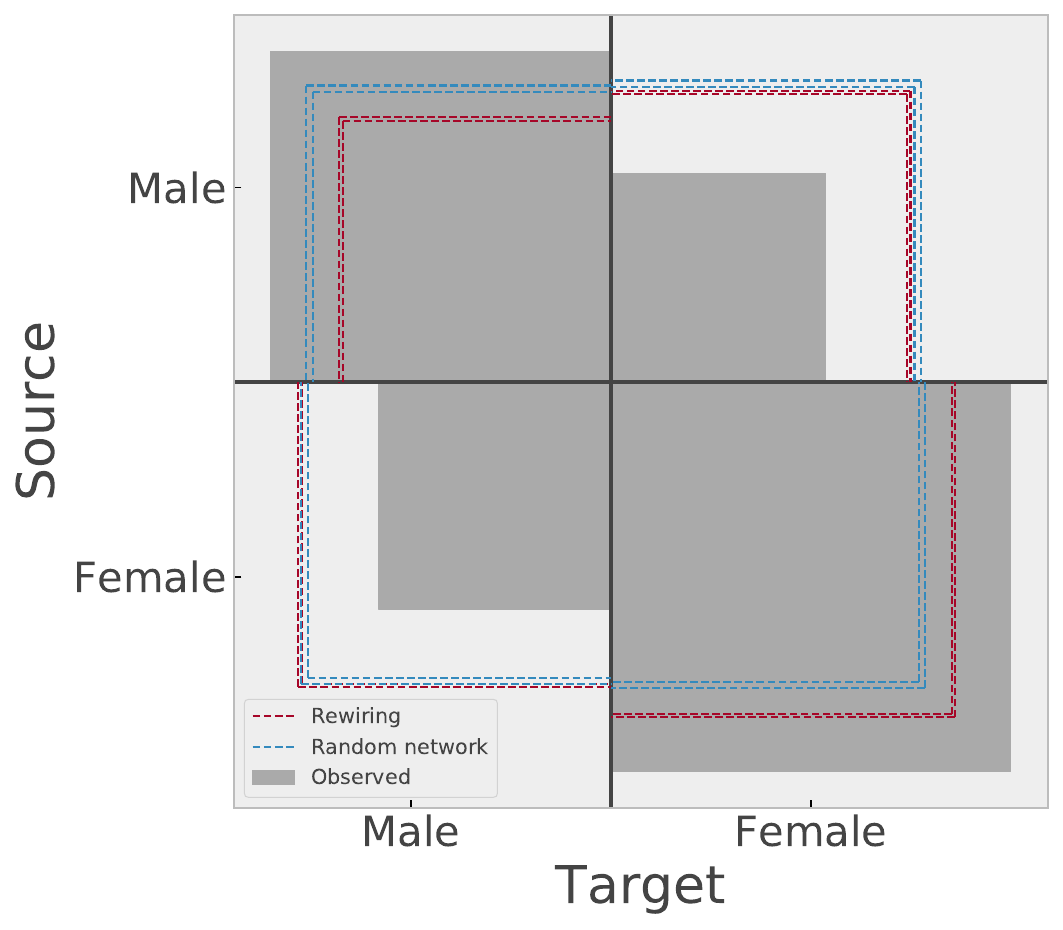}
\caption{{\bf Following patterns by sex.} Representation of the number of links between sources (followers) and targets (followees) by sex. Areas are proportional to the number of links in each entry, where the shaded areas depict the observed links, and the area enclosed by the red and blue dashed lines stands for, respectively, the rewired and random directed network null models, with two lines for each colour delimiting a standard deviation from the average behaviour.}
\label{figs:fig3}
\end{figure}

We used the individuals' metadata, including the sex, country and sport, to quantify the observed homophily in the network (\emph{i.e.}, the likelihood of following peers with similar features, known as assortativity in the context of network science). In general, in terms of the observed links, 78.8\% of them connected two individuals winning medals in the same sport, 74.4\% of them occurred between representatives of the same country and 73.3\% linked athletes with the same sex, indicating the presence of highly assortative patterns in the network.

%Fig. 4. Country assortativity
\begin{figure*}
\includegraphics[width=\textwidth]{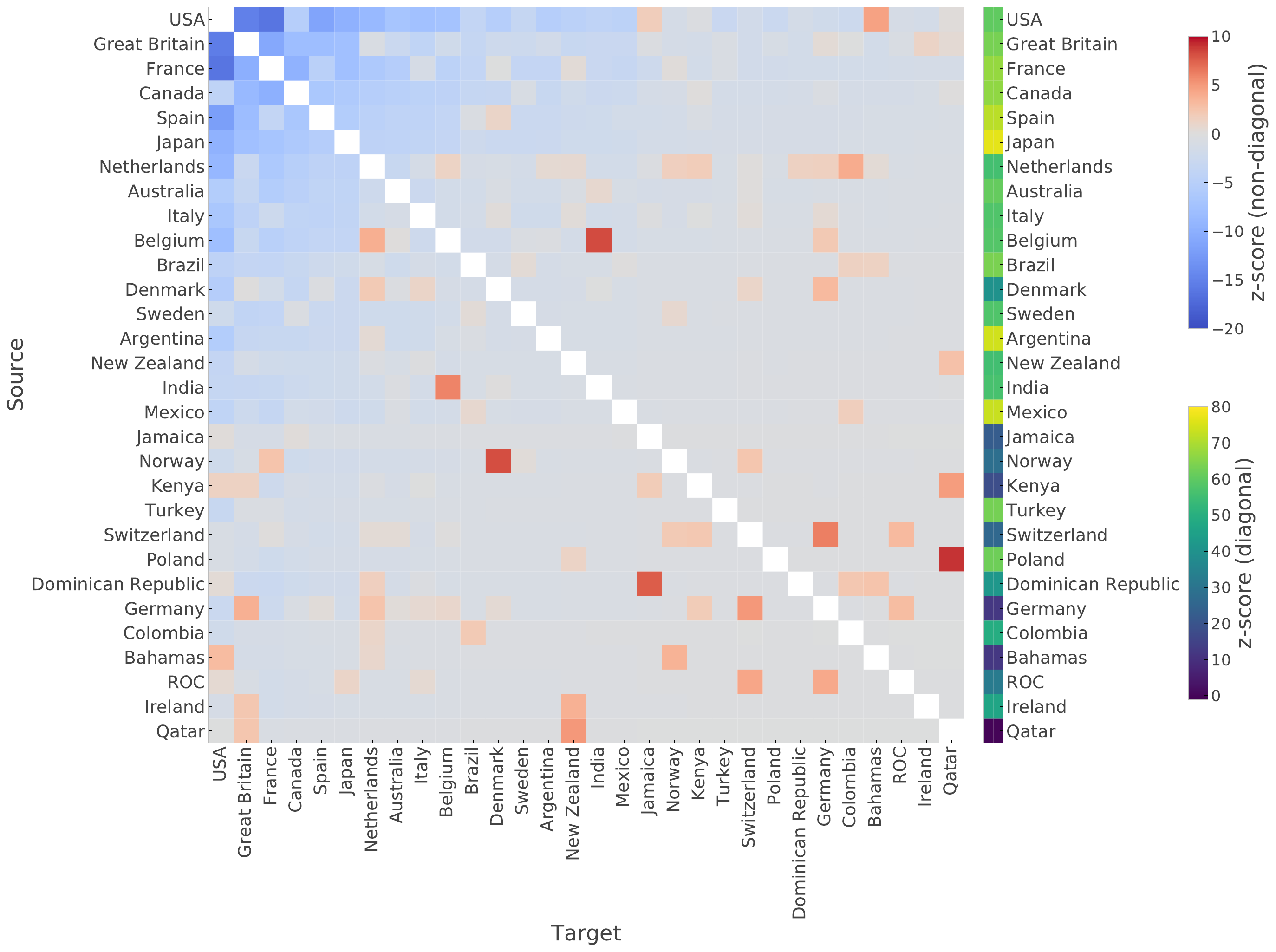}
\caption{{\bf Following patterns by country.} Matrix representing how many following links occurred between different pairs of countries, represented as the $z$-score of the observed links with a reference to what would be expected in a random scenario, which was given by 1000 realizations of the network rewiring. The diagonal and non-diagonal terms span across different scales, so the former are represented in the right bar.}
\label{figs:fig4}
\end{figure*}

Regarding the sex feature, we measured the total in- and out-degrees of each sex, obtaining $k_{\text{in}}^M$, $k_{\text{in}}^F$, $k_{\text{out}}^M$ and $k_{\text{out}}^F$ by adding all the individual in- and out-degrees according to the athlete's sex. Then, we used the ratio between these in- and out-degrees per sex to estimate the direction of the disassortative links in the network, obtaining $k_{\text{in}}^{M}/k_{\text{out}}^M = 1.05$ and $k_{\text{in}}^{F}/k_{\text{out}}^F = 0.96$. These results indicated that male athletes had more followers than followees, implying that the disassortative links where female athletes followed male athletes were more likely than otherwise, which was confirmed by a matrix visualization of the observed links considering the source (follower) and target's (followee) sexes (Fig. \ref{figs:fig3}). First, this matrix highlighted the assortative patterns through the stronger diagonal terms in contrast to the non-diagonal entries (shaded areas on Fig. \ref{figs:fig3} represented the number of observed links in each entry). To understand the significance of this pattern and the non-diagonal terms, we proposed two reference null models. First, we rewired the edges by considering pairs of links and swapping their targets (\emph{e.g.}, if we observed $a\rightarrow b$ and $c\rightarrow d$, a rewire between these links would lead to $a\rightarrow d$ and $c\rightarrow b$); this reference model would keep fixed individual properties such as the sex, $k_{\text{in}}$ and $k_{\text{out}}$. Second, we created a random directed network with the same number of nodes, edges and number of female/male athletes. Both reference models underestimated the diagonal terms and overestimated the non-diagonal cases. In the diagonal, the rewiring model displayed a higher number of male $\rightarrow$ male links than the random network model; while we observed the opposite behaviour for the female $\rightarrow$ female links. We understood this behaviour taking into account that the number of links in the random network displayed minor variations across the different matrix entries as the number of male and female athletes was similar (475 male and 489 female), while for the rewired network, the expected number of male $\rightarrow$ male links would be proportional to $k_{\text{in}}^M k_{\text{out}}^M$ (and similarly for female $\rightarrow$ female); thus, as $k_{\text{in}}^F k_{\text{out}}^F$ was 60\% larger than $k_{\text{in}}^M k_{\text{out}}^M$, the rewiring provided a higher estimate of female $\rightarrow$ female links than the random network model, while this estimate was lower than the random network model for male $\rightarrow$ male links.

Focusing on the athlete's countries, we created a directed weighted network described by the weights $w_{ij}$ that indicated how many following links existed between athletes from countries $i$ (followers) and $j$ (followees). Considering what we observed in the sex analysis, we chose the rewiring null model as our reference, as this model kept the features (sex, country and sport), together with the in- and out-degrees of the nodes. We performed multiple realizations of the rewiring of the network at the individual athlete level, obtaining from each realization $r$ the weights $w_{ij}^r$. This allowed us to quantitatively compare the observed weights with the null model through a $z$-score:
\begin{equation}
z_{ij}=\frac{w_{ij}-\langle w_{ij}^r \rangle}{\sigma(\langle w_{ij}^r \rangle)}
\end{equation}
where $\langle w_{ij}^r \rangle$ was the average weight over the multiple rewiring realizations and $\sigma$ was its standard deviation. The representation of these $z_{ij}$ for the 30 countries with highest in-degree, as a matrix connecting source to target countries, revealed different scales for the diagonal and non-diagonal entries (Fig. \ref{figs:fig4}). To better visualize both terms, we extracted the diagonal term, observing generally large values. Particularly, all the countries had a diagonal $z$-score $z_D > 10$ except for Qatar ($z_D = - 0.11$), indicating high homophily at country level. The highest diagonal scores were those of Japan ($z=76.75$), Argentina ($z=74.12$) and Mexico ($z=73.23$). The generic result in the non-diagonal entries was the negative $z$-scores for the most popular countries, which we associate with the increase of the in-degrees due to the diagonal interaction, leading to a defect of interactions in the non-diagonal entries, together with a low $z-$score pattern for the rest of countries (less diagonal, more random interactions). However, we observed a few with high $z$-score, including Poland $\rightarrow$ Qatar ($z=9.04$), Belgium $\rightarrow$ India ($z=8.24$), Norway $\rightarrow$ Denmark ($z=8.18$) and Dominican Republic $\rightarrow$ Jamaica ($z=7.65$). Although these values are high $z$-scores, they did not reach the typical values of the diagonal scores.

\subsection{Modelling interactions}
%Fig. 5. Model
\begin{figure*}
\includegraphics[width=\textwidth]{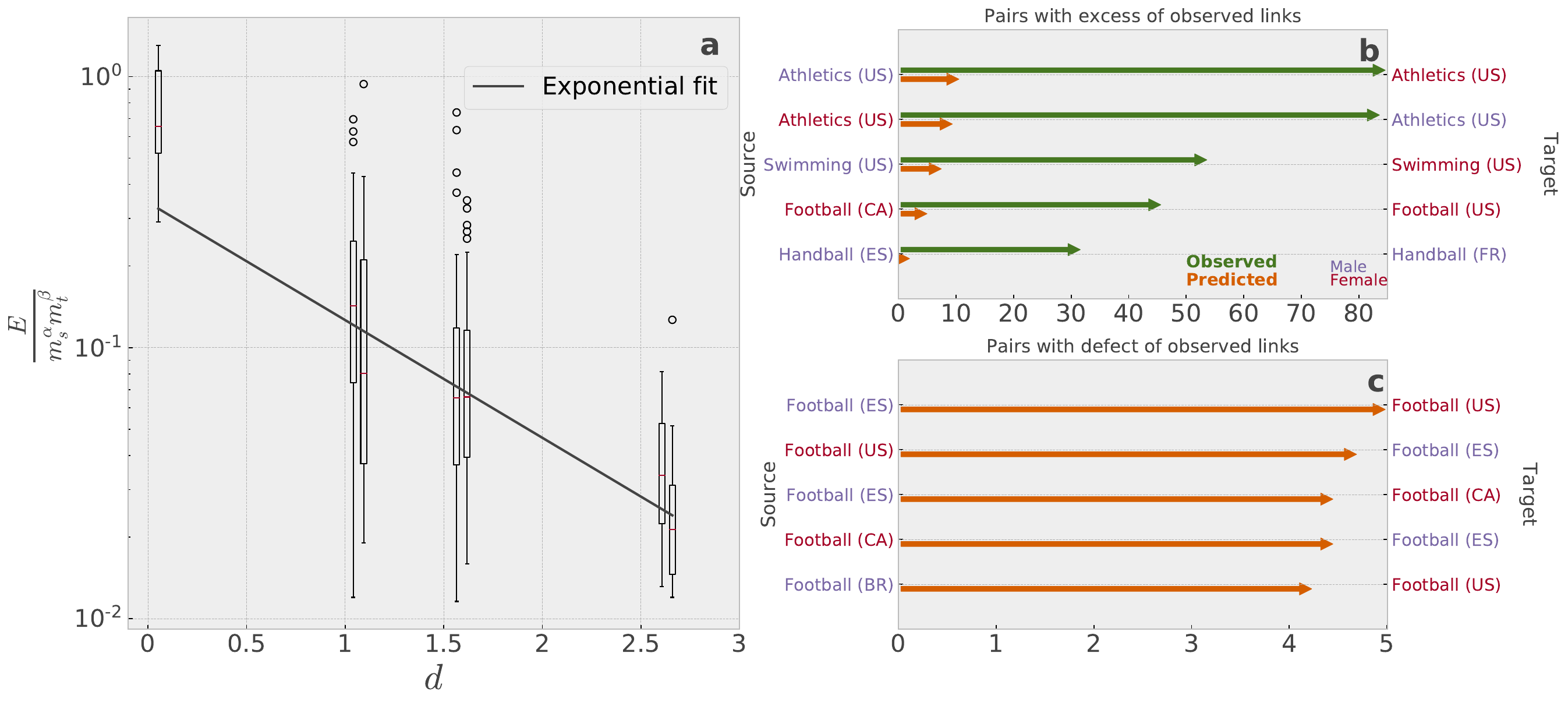}
\caption{{\bf Modelling interactions between different groups. a,} Number of following links $E$, normalized by the scaled source (follower) and target masses, $m_s^{\alpha}$ and $m_t^{\beta}$, respectively, as a function of the distance between genotypes $d$, computed with the estimated weights from a linear regression in semi-log scale. The data is shown for the seven possible combinations of same or different sex, country and sport, with the solid line representing an exponential regression, with the horizontal red line depicting the median of the data for each $d$, the 25-75 percentiles represented in the boxes, the whiskers representing the 25 (75) percentile minus (plus) 1.5 the interquartile range, and the isolated dots showing the outliers. {\bf b,} Top-5 edges in the genotype network with an excess of weight in the observed pattern that the model cannot explain; these represented links obtained a $p$-value (see Methods) of 0 after $10^6$ stochastic realizations of the model, and they are sorted by the observed weight. {\bf c,} Top-5 edges in the genotype network with a defect of weight, sorted by decreasing $p$-value. The $x$-axis in {\bf b} and {\bf c} represents the link weight, which is observed in the data (green arrows) or predicted by the model (orange arrows). The $y$-axis labels report the source and target's sex and country, with the colour standing for their sex. Country abbreviations are the following: United States (US), Canada (CA), Spain (ES), France (FR) and Brazil (BR).}
\label{figs:fig5}
\end{figure*}

As we frequently observed links between individual athletes with similar features (sex, country, sport), we developed a macroscopic model to characterise the relative importance of such features in the homophilic patterns that our network displayed. This model was based on an auxiliary macroscopic network where the nodes were genotypes, introduced as the observed unique combinations of the triad (sex, country, sport). Each genotype $i$ had a mass $m_i$ that was the number of athletes with that genotype and at least one link in the network. The macroscopic network was created considering the genotypes with $m\geq5$, did not include self-loops (\emph{i.e.}, following links within the same genotype), was directed (from followee to follower genotypes) and weighted, where the weights were the number $E_{ij}$ of athletes with genotype $i$ following peers with genotype $j$, representing 24.7\% of the observed links between individual athletes. We proposed a gravity-like approach to model the interactions in this macroscopic network. Gravity models mimic Newton’s law of gravitation \cite{zipf1946p}, assuming that the flow between two nodes in the network scales with the product of their masses and decays with the distance between the nodes, and have been applied to traffic and human mobility flows in many empirical contexts \cite{lenormand2016systematic,kaluza2010complex}. In our approach, we considered that $E_{ij}$ depended on the masses of genotypes $i$ (follower) and $j$ (followee) and the distance $d_{ij}$ that reflected the difference between these genotypes:
\begin{equation}
E_{ij} = \frac{m_i^{\alpha} m_j^{\beta}}{f(d_{ij})}
\label{eqgravity}
\end{equation}
where $\alpha$ and $\beta$ were the scaling exponents for the masses of the genotypes that acted as, respectively, follower and followee, and $f(d_{ij})$ was a monotonous function of the feature-based distance between genotypes. 

First, we estimated the values of $\alpha$ and $\beta$ by selecting sub-samples of this macroscopic network with a fixed distance. Particularly, we checked the seven different combinations of similar/different sex, country and sport (\emph{i.e.}, as there are 2 possibilities, similar/different, and 3 features, there would be 8 combinations, but the combination with the same features, the self-loop, was not considered in our network). For each of these combinations, where all the links would have the same distance, we made the linear regression of $\log E_{ij} = \alpha \log m_i + \beta \log m_j + K$, with $K$  being a constant. After fitting this equation for each of the distance combinations, we selected the values of $\alpha$ and $\beta$ from the combination with largest correlation coefficient, which was $r=0.55$ for the case of genotypes having the same country, the same sport and a different sex, leading to $\alpha=0.50$ and $\beta=1.07$.

Second, we estimated the relative importance of each feature in the distance between genotypes. Particularly, we introduced distance $d_{ij}$ between genotypes $i$ and $j$ as
\begin{equation}
d_{ij} = w_{\text{sx}}d_{ij}^{\text{sx}}+w_{\text{c}}d_{ij}^{\text{c}}+w_{\text{sp}}d_{ij}^{\text{sp}}
\label{eqw}
\end{equation}
where $w_k$ was the weight of feature $k$ on the distance, sx, c and sp standing for, respectively, sex, country and sport, and $d_{ij}^{k}=0$ if the feature $k$ was the same for the genotypes $i$ and $j$ and 1 otherwise. We fitted the normalized weight $E_{ij}/(m_i^{\alpha} m_j^{\beta})$ as a function of inverse powers of $d_{ij}$, observing that the correlation coefficient increased with the exponent of $d_{ij}$, which suggested fitting an exponentially decreasing function of the distance. Consequently, we fitted $\log \frac{m_i^{\alpha} m_j^{\beta}}{E_{ij}} = w_{\text{sx}}d_{ij}^{\text{sx}}+w_{\text{c}}d_{ij}^{\text{c}}+w_{\text{sp}} d_{ij}^{\text{sp}} + K'$, with $K'$ being a constant, to obtain $w_{\text{sx}}=0.053$, $w_{\text{c}}=1.04$ and $w_{\text{sp}}=1.57$, with a correlation coefficient $r=0.32$ (Fig. \ref{figs:fig5}a). We measured the robustness of these weight estimates through multiple fits, where we removed iteratively the data associated to one of the seven possible combinations of same/different sex, country and sport (\emph{i.e.}, each iteration considered all the data except for that associated to one combination), leading to the average values over different iterations $\langle w_{\text{sx}} \rangle= 0.07 \pm 0.13$, $\langle w_{\text{c}} \rangle= 1.08 \pm 0.19$, and $\langle w_{\text{sp}} \rangle= 1.58 \pm 0.16$. 

With the estimated weights from the model, we forecasted the links between every pair of genotypes with a mass higher or equal than 5 individuals. Specifically, we distributed the empirically observed $E=\sum_{i\neq j} E_{ij}=1808$ links among these pairs by computing the probability of a link between an individual with genotype $i$ (source) and another with genotype $j$ (target):
\begin{equation}
P_{ij} = E \frac{m_i^{\alpha} m_j^{\beta} \exp(-d_{ij})}{\sum_{i\neq j} m_i^{\alpha} m_j^{\beta} \exp(-d_{ij})}
\label{eqpij}
\end{equation}

Each link of the possible $m_i m_j$ links occurred with this probability. After a realization, if the number of links $i\rightarrow j$ was higher than the number of links of individuals with genotype $i$ following those with genotype $j$, we increased the $p$-value associated with that connection (see Methods). Consequently, high $p$-values were associated with a defect of observed microscopic links with respect to the model, and low $p$-values with an excess of observed links. After $10^6$ realizations, there were several macroscopic links with $p=0$ (excess of observed links), so we ranked them according to the number of observed microscopic links (Fig. \ref{figs:fig5}b). The first three were associated with connections between athletes from the United States, practising the same sport (athletics and swimming), but with different sexes, the fourth connected female athletes playing football from Canada and the United States, while the fifth linked the Spanish male handball players to the French, with the same sex and sport. The macroscopic links with the highest $p$-values did not display any link in the observed data, and the ranking according to the $p$-values matched the ranking according to the number of predicted links. Particularly, the top-5 links with highest $p$-values connected football players with different sex and from different countries (Fig. \ref{figs:fig5}c). Note that, for the case of football, links between players from the same country would not be possible, as the medal-awarded countries were different for male and female competitions. For these links, while there were no observed links, the number of predicted links ranged between 4 and 5, leading to high $p$-values due to the unlikelihood of a realization leading to no links between those genotypes.

\section{Discussion}

Complex networks are powerful tools to understand and simulate interacting dynamics among populations. Consequently, the inference of such networks is useful to extract the features underlying the temporal evolution of systems composed of multiple coupled particles, which for example can be agents in a social system. In this context, there are phenomena that strike quantitative social scientists, such as why some discussion topics emerge at specific moments. For example, considering the population of Olympic athletes, there have been multiple attempts to discuss on the mental health problems associated with sport, but this topic earned relevance on the Olympic Games of Tokyo 2020 after Simone Biles decided not to compete in several disciplines. The network that we inferred may explain the emergence of this topic during this event, considering the high centrality (according to the in-degree) of this athlete on the network. This particular example highlights the role that influential spreaders play in the spread of information on top of complex social networks, and shows that the analysis of the \emph{rich-club} subset of Olympic medallists allows identifying key nodes that are able to reach a broad audience and therefore influence the general opinion in a short time.

Complementing the key role of influential spreaders, dynamical models of social phenomena frequently consider homophily as a driver of interaction, but generally the lack of data leads to assumptions such as equal weighting across different features \cite{axelrod1997complexity,klemm2003nonequilibrium}. Our model of social interaction used a macroscopic network to connect homogeneous groups, allowing us to quantify the importance of the features in the choice of followees, which can be key for informing dynamical models. Although the microscopic sex interactions displayed a high assortativity (Fig. \ref{figs:fig3}), the sex had a negligible role in the macroscopic network, with a contribution to the distance that was two orders of magnitude lower than the country and the sport, and even compatible with the zero-weight scenario in our robustness test; this was highlighted in the representation of the model regression and the data, where the sex differences implied minor changes in the observed data (Fig. \ref{figs:fig5}a). In contrast, the sport was the most influential feature, having a weight that was 50\% higher than that of the country. We also estimated the scaling exponents of the source and target masses to be, respectively, $\alpha=0.50$ and $\beta=1.07$. While the $\beta$ value indicated a linear behaviour (\emph{i.e.}, the probability of creating a link to a group is proportional to the group size), $\alpha$ described a sub-linear process, where genotypes with large sizes had a tendency to follow less frequently individuals with other genotypes. This behaviour suggests that when groups become bigger, they are more likely to create internal links rather than establishing links towards different genotypes, agreeing with the high assortativity that we observed in the microscopic network. However, we acknowledge that other factors such as the prestige of the awarded medals, international or even personal relations may influence the dynamics of this network creation, which highlight the presence of other relevant dimensions that are neglected in our approach.

Here we have focused on the Tokyo 2020 snapshot of the follower-followee network of Olympic medallists, as this represents the underlying structure of the information flows among these athletes on Twitter at that time. %, although our analysis presents some limitations. 
In addition to the structure, a more detailed approach to the information flows should consider the algorithm that decides what posts from a user's followees are shown to them, which could be the reason for collective attention being dominated by a few individuals \cite{morales2014efficiency}. Moreover, other works analysing Twitter data have considered multiple kinds of interactions (and not only the follower-followee scenario), highlighting its multilayer structure, for example through the query of mentions (posts where a user tagged another) or retweets (posts by another user that an individual shares with his followers) \cite{borondo2015multiple}. Adding the temporal axis to our analysis may also capture important aspects of the system, and a comparison between our dataset and the one of the incoming Olympic Games in Paris 2024 would reveal more nuanced trends and add more statistical evidence to the present findings. 

Despite the aforementioned limitations, this work advances the understanding of social dynamics by applying quantitative modelling techniques to empirical network data that is available thanks to the high public exposure and global impact of high-performance athletes. Besides the advantages in data availability, this particular system has some interesting universal properties, which are revealed by our modelling approach. We have shown that the complex interaction patterns are strongly driven by metadata information (sex, nationality and sport) which are in fact the most relevant individual features in many other social dynamics beyond this particular context (if one naturally interprets the sport feature as the occupation sector or speciality of the individuals). Our gravity-like framework, using the macroscopic network of genotypes and the concept of feature-based distances between genotypes, can therefore be applied to other contexts, such as following networks of politicians in institutions at different scales, potentially revealing and quantifying subtle information flows, assortative biases and polarization effects that may be currently underestimated. On the other hand, the presented dataset offers an interesting metadata-enriched complex network, which can be leveraged to design and test novel methodologies and hypothesis from multidisciplinary perspectives. Taken all together, our analysis of the following network of Olympic medallists provides a suitable proxy to better understand the fingerprints of relevant network properties, such as popularity, reciprocity and (metadata) assortative patterns, that emerge in many large and interconnected socio-cultural systems at different scales.

\section{Methods}
{\bf Twitter network.} We created a database with the Twitter usernames of 1052 medallists in the Tokyo 2020 Olympic Games (which took place on 2021), according to the information available in Wikipedia \footnote{\url{https://en.wikipedia.org/wiki/List_of_2020_Summer_Olympics_medal_winners}}, including only those with an open Twitter account and specifying their name, Twitter username, sex, sport, country and which medal(s) they got. Introducing this information in the Twitter API, we checked the accounts that each medallist with a Twitter account followed, and created a database with the accounts that belonged to other medallists, leading to the creation of a complex directed network describing who followed whom. This network illustrates the outcome of our queries to the Twitter API on January 2023.

{\bf Linear regression.} We performed the multiple linear regression tests using the LinearRegression function of the \emph{sklearn} package in Python.

{\bf $p$-value estimation to compare the model with empirical observations.} Considering the model fitted model parameters, represented in Eqs. (\ref{eqgravity},\ref{eqw}), the probability of an individual with genotype $i$ following another with genotype $j$ was given by Eq. (\ref{eqpij}). We performed $R=10^6$ realizations of this link random network generation. From each realization $r$ of this microscopic network, we extracted the macroscopic network and compared, for each pair of genotypes $i$ (source) and $j$ (target), the number of observed links $E_{ij}$ and the number of predicted links on the realization $E^{r,\text{pred}}_{ij}$, increasing the $p$-value $p_{ij}$ in 1/$R$ when $E^{r,\text{pred}}_{ij}>E_{ij}$, that is, when the model predicted an excess of microscopic links with respect to the observed number of links. Consequently, high $p$-values were associated with systematic excess of predicted links, while low $p$-values likely represented a defect.

\section{Data and availability}
Authors will make all the data (network and individual metadata) and codes available on a public repository upon acceptance of the manuscript.

\section{Author contributions}
Conceptualization: J.P.R. Data: J.P.R. Software: J.P.R. Analysis: J.P.R. and L.A.F. Visualization: J.P.R. Writing -- Original draft: J.P.R. and L.A.F.

\begin{acknowledgments}
J.P.R. was supported by the Juan de la Cierva Formación program (Reference No. FJC2019-040622-I) funded by the Spanish Ministry of Science and Innovation, acknowledges funding from the Spanish Research Agency MCIN/AEI/ 10.13039/501100011033 via project MISLAND (PID2020-114324GB-C22), and by Govern de les Illes Balears through the Vicenç Mut program. L.A.-F. acknowledges funding from DYNDEEP (EUR2021-122007) from the Agencia Estatal de Investigaci\' on MCIN/AEI/10.13039/501100011033. This research was supported by María de Maeztu Excellence Unit 2023-2027 Ref. CEX2021-001164-M, funded by MCIN/AEI/ 10.13039/501100011033.
\end{acknowledgments}

\bibliographystyle{IEEEtran}
% Generated by IEEEtran.bst, version: 1.14 (2015/08/26)

\end{document}